\documentclass[12pt]{article}    
\usepackage[ansinew]{inputenc} 
\usepackage{hyperref}
\usepackage{graphics}
\usepackage{epsfig}
\usepackage{epstopdf}
\usepackage{graphicx} 
\usepackage{amssymb}  
\usepackage{amsmath}
\usepackage{multirow} 
\usepackage{enumitem} 
\usepackage{color}     
\usepackage{indentfirst}  
\usepackage{tabularx,booktabs}
\usepackage{caption} 
\usepackage{appendix}
\renewcommand\appendix{\section*{Appendix}}
\usepackage[gen]{eurosym}
\usepackage[scientific-notation=true]{siunitx}
\usepackage{epstopdf}

\usepackage[all]{xy}
\usepackage{bm}

\usepackage{lscape}
\usepackage{afterpage}

\begin{document}
\newtheorem{definition}{Definition}
\newtheorem{theorem}{Theorem}
\newtheorem{example}{Example}
\newtheorem{corollary}{Corollary}
\newtheorem{lemma}{Lemma}
\newtheorem{proposition}{Proposition}
\newtheorem{remark}{Remark}
\newenvironment{proof}{{\bf Proof:\ \ }}{\qed}
\newcommand{\qed}{\rule{0.5em}{1.5ex}}
\newcommand{\bfg}[1]{\mbox{\boldmath $#1$\unboldmath}}

\begin{center}

\section*{Inequality in the use frequency of patent technology codes}

\vskip 0.2in {\sc \bf Jos\'e Alejandro Mendoza$^a$
Faustino Prieto$^b$
\footnote{Corresponding author. E-mail addresses: faustino.prieto@unican.es.}, 
Jos\'e Mar\'{\i}a Sarabia$^c$
\vskip 0.2in

{\small\it
$^a$University of Cantabria, Santander, Spain\\
$^b$Department of Economics, University of Cantabria, Santander, Spain\\
$^c$Department of Quantitative Methods, CUNEF University, Madrid, Spain\\}
}
\end{center}

\begin{abstract}
Technology codes are assigned to each patent for classification purposes and to identify the components of its novelty. Not all the technology codes are used with the same frequency - if we study the use frequency of codes in a year, we can find predominant technologies used in many patents and technology codes not so frequent as part of a patent. In this paper, we measure that inequality in the use frequency of patent technology codes. First, we analyze the total inequality in that use frequency considering the patent applications filed under the Patent Co-operation Treaty at international phase, with the European Patent Office as designated office, in the period 1977-2018, on a yearly basis. Then, we analyze the decomposition of that inequality by grouping the technology codes by productive economic activities.  We show that total inequality had an initial period of growth followed by a phase of relative stabilization, and that it tends to be persistently high. We also show that total inequality was mainly driven by inequality within productive economic activities, with a low contribution of the between-activities component.\\

\noindent {\bf Key Words}: Inequality; innovation; technology codes; patents

\end{abstract}

\section{Introduction}

Technology codes are used to classify a new patent into specific technical categories and identify the components of the invention's novelty \cite{You15}. They can also be used, for example, to analyze the productivity of innovation \cite{Str10}, to 
track and characterize technological change \cite{Str12}, to identify the sources of technological novelty \cite{Str15}, to determine the technology development trends \cite{Cha17}, to extract the technological inventions trends by group of applicants \cite{Cha19},  or to track the technological composition of industries \cite{Gol20}.

Whenever an invention is made, and its inventor decide to patent it, one or more technological codes are assigned to the patent, reflecting a process of innovation based mainly in the knowledge of the available technologies, the combination of those technologies, and occasionally the creation of new technologies \cite{Wei98,Art09}. However, not all the technology codes are used with the same frequency \cite{Hua16,Smo16}. On the one hand, we can find mainstream technologies being part of many patents, that could be associated with technological areas with larger potential market, and on the other hand, there are uncommon technologies being part of unusual technological combinations \cite{Har19}.

The aim of this paper is to measure the inequality in the use frequency of patent technology codes.

Inequality is a key concept in economics \cite{Sen97}, with a large proportion of the literature dedicated to income inequality \cite{Pik03,Sar19}, but also focused on wealth inequality \cite{Cag08,Zuc19}, inequality in education \cite{Til87,Jor17,Coc20}, or inequality in productivity \cite{Ezc07}, among others,  and that can be extended to other variables. In particular, within the framework of the economics of innovation, 
the efforts have been focused, among others, on the impact of innovation in economic growth \cite{Adr19},  the returns to innovations \cite{Mar05}, the technological specialization in metropolitan areas \cite{OHu11} or in countries \cite{Man21}, or the role of institutions \cite{Biu20}.  However, to our knowledge, there are no previous efforts to measure the inequality in the use frequency of technology codes among the different patents.

For our analysis, we considered the OECD REGPAT patent database, in which the International Patent Classification (IPC) and the Cooperative Patent Classification (CPC) technology codes are available. We selected the patent applications filed under the Patent Co-operation Treaty (PCT) at international phase, with the European Patent Office (EPO) as designated office, with priority date in the period 1977-2018, and then, we measured the total inequality in use frequency per year of those technology codes. For that, we used two well-known inequality meausures: the Gini index and the Theil index. Finally, based on the decomposition of the Theil index, and by using the correspondence between IPC technology codes and NACE (European classification of economic activities) divisions, we studied the decomposition of that total inequality by productive economic activities and calculated the contribution of the within-activity and between-activity components in the total inequality previosly measured.

The rest of this paper is organized as follows: in Section \ref{sec:2}, we describe the dataset and methodology used; in Section \ref{sec:3}, we present the results obtained; finally, conclusions are given in Section \ref{sec:4}.

\section{Data and Methodology}
\label{sec:2}

\subsection{Patent database}\label{subsection21}

For this study, we used patent information obtained from the OECD REGPAT database, July 2021 edition, (see \cite{Mar08}, for a detailed presentation of that database), published by the Organisation for Economic Co-operation and Development, and available by filling a form on the OECD website \cite{Oec21}. 

We considered the patent applications filed under the Patent Co-operation Treaty (PCT) at international phase, with the European Patent Office (EPO) as designated office, and with priority date over the period 1977-2018. We used the priority date as the reference year for each patent, following the recommendation of the OECD 
(see \cite{Oec09}, p.62) as the closest date available to when that invention was conceived. And taking into account that a patent application is usually published after 18 months, we selected that period of 42 years, from 1977 to 2018.

We analyzed the use frequency per year of patent technology codes, for both IPC and CPC codes (corresponding to the International Patent Classification and the Cooperative Patent Classification systems, respectively). For that, we obtained the information about IPC codes directly from the file ``202107\_PCT\_IPC.txt" file, that gives for each patent: the PCT publication number (pct\_nbr), the priority year (prio\_year), the EPO filling year (app\_year) and its IPC codes. Then, we obtained the information about CPC codes (for the same patents) using additionally the ``202107\_PCT\_App\_reg.txt" and ``202107\_CPC\_Classes.txt" files, and joining the three files by using the fields pct\_nbr and appln\_id (EPO worldwide patent statistical database, PATSTAT, application identifier).

\begin{figure}[t]
\centering
\includegraphics[width=1\textwidth]{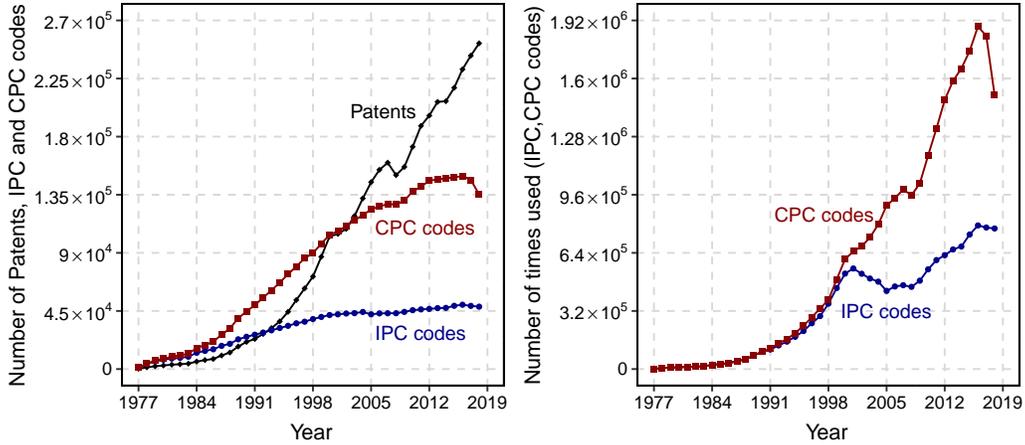}
\caption{Number of PCT(EPO) patents, and number of different IPC and CPC codes per year (on the left). Total number of times a year that those IPC and CPC codes were used (on the right). Period: 1977-2018.} \label{fig.01}
\end{figure}

In summary, our dataset was composed of 3,786,795 patents, with 72,607 different IPC codes that were used 13,615,554 times in total over the period 1977-2018, and also with 234,782 different CPC codes that were used 25,227,358 in all those patents over the same period of time. Figure \ref{fig.01} shows, on the left, the number of patents, and the number of different IPC and CPC codes per year, and on the right, the total number of times a year that those IPC and CPC codes were used.

\subsection{Use frequency of IPC and CPC codes}

Our random variables of interest were the use frequency a year of IPC codes ($X_1$)  and the use frequency a year of CPC codes ($X_2$). It can be noted that an analysis of the period 1977-2018, all together, would group codes with different age, so we conducted our analysis on a yearly basis, by grouping codes used in the same year. As a reference, Table \ref{tab.01} shows the frequency tables of $X_1$ and $X_2$ from our dataset, corresponding to the year 1978. It can be seen that in 1978, there were 3260 IPC codes and 3791 CPC codes that were used just once (in only one patent), and there were one IPC code used 18 times and another CPC code used in 15 different patents.

\begin{figure}[htp]
\centering
\includegraphics[width=1\textwidth]{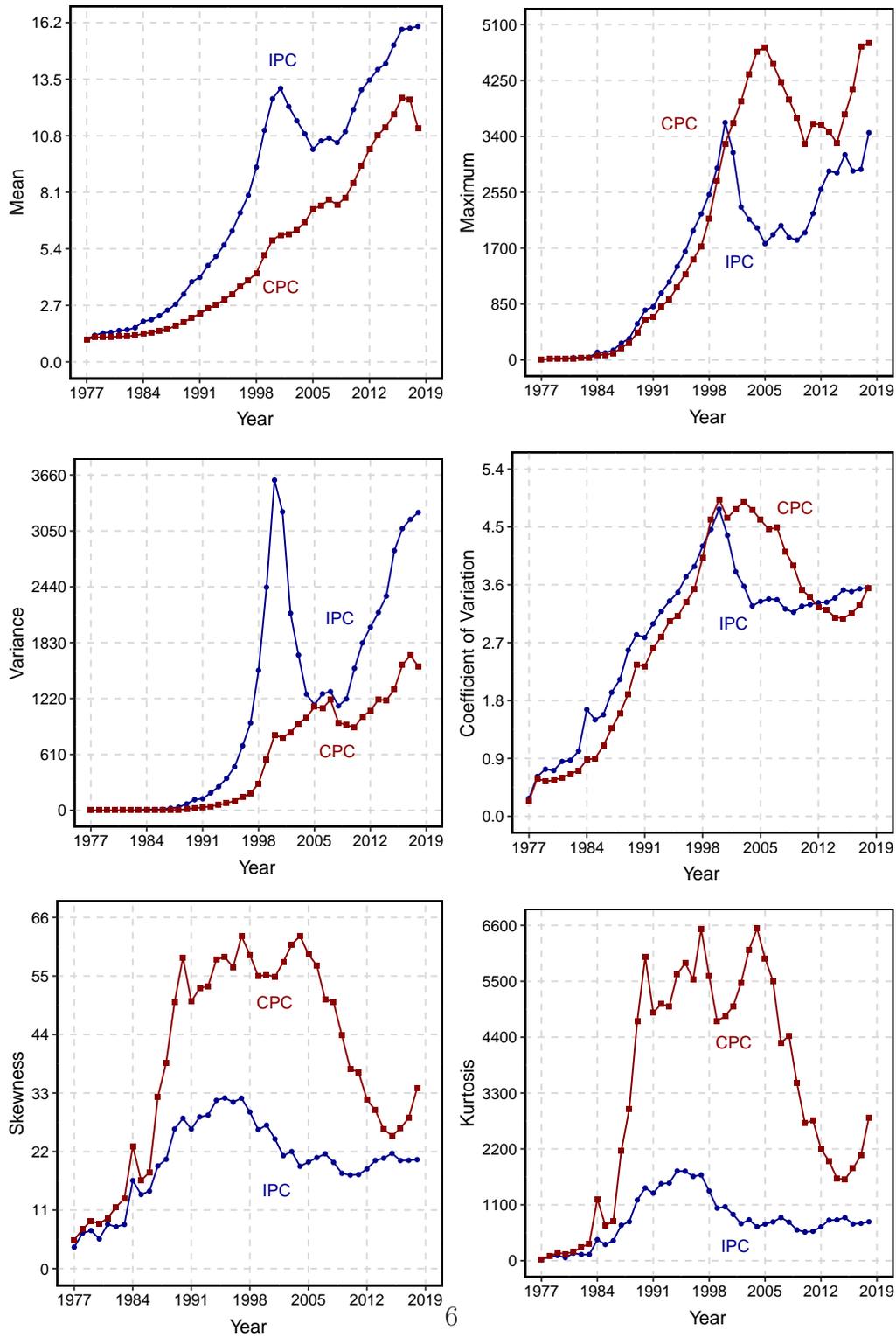}
\caption{Some relevant information about the datasets considered.} \label{fig.02}
\end{figure}

Figure \ref{fig.02} shows the main empirical characteristics of both variables of interest: the mean, the maximum, the variance, the coefficient of variation, the skewness and the kurtosis of the use frequency per year, for IPC and CPC technology codes. For example, it can be noted that IPC codes were used more than CPC codes on average, in all the years considered, reaching a mean value of use frequency in 2018, for IPC and CPC codes, of approximately 16 times and 11 times respectively. Additionally, both variables exhibited positive skewness and high kurtosis, with greater values in both measures in the case of CPC codes, and the values of the coefficient of variation (relative dispersion of the data) were not very different between IPC and CPC codes.

\begin{table}[t]\footnotesize
\setlength{\tabcolsep}{8pt}
\caption{Frequency tables (for IPC and CPC codes) corresponding to the year 1978.}
\begin{tabularx}{\textwidth}{ccp{0.6cm}cc} 
\toprule
\multicolumn{2}{l}{IPC codes. Year: 1978}& & 
\multicolumn{2}{l}{CPC codes. Year: 1978} \\ 
 \cmidrule{1-2} \cmidrule{4-5}
$x_{1i}$& $n_{1i}$& &
$x_{2i}$& $n_{2i}$\\ 
(use frequency)&(number of codes)& &
(use frequency)&(number of codes)\\ 
 \cmidrule{1-2} \cmidrule{4-5}
1 & 3260 & &   1 & 3791 \\
2   & 521 & &   2 &   338 \\
3   & 116 & &   3 &     55 \\
4   &   48 & &   4 &     54 \\
5   &   19 & &   5 &     14 \\
6   &     4 & &   6 &       6 \\
7   &     2 & &   7 &       4 \\
8   &     3 & &   8 &       2 \\
9   &     3 & &   9 &       2 \\
12 &     2 & & 11 &       1 \\
18 &     1 & & 13 &       1  \\
     &        & & 15 &        1  \\ \bottomrule
\end{tabularx}
\label{tab.01}
\end{table}

\subsection{Inequality measures}

For our analysis, we used the Gini and Theil indexes \cite{Arn18}, two of the most commonly used inequality measures.  
\begin{itemize}
\item The Gini index was calculated by using the well-known Brown formula \cite{Bro94}, given in its original notation by:
\begin{equation}\label{Eq.01}
G=1-\sum_{i=0}^{k-1}(Y_{i+1}+Y_i)(X_{i+1}-X_i)
\end{equation}
where $Y$ is the cumulative proportion of use frequency in a year, and $X$ is the cumulative proportion of codes, over $k$ values of use frequency ordered from the codes used just once (used in only one patent) to the codes most often used in the year considered.
The Gini index is related with the Lorenz curve as double the area between the Lorenz curve and the line of perfect equality. 
\item The Theil index \cite{All78,Sho80} was calculated as follows:
\begin{equation}\label{Eq.02}
T({\bm y},n)=\displaystyle\frac{1}{n}\sum_{i=1}^n\left(\displaystyle\frac{y_i}{\mu}\right)
\log\left(\displaystyle\frac{y_i}{\mu}\right)
\end{equation}
where ${\bm y}=(y_1,\dots,y_i,\dots,y_n)$ is the use frequency a year vector for a population of $n$ technology codes (see figure \ref{fig.01}, on the left), $\mu$ is the mean value of the use frequency per year (see figure \ref{fig.02}), and where natural logarithms are used.
\end{itemize}

\subsection{Inequality decomposition by NACE divisions}

NACE (Nomenclature statistique des activit\'es \'economiques dans la Communaut\'e europ\'eenne) is a European classification scheme of productive economic activities \cite{Eur08}. It has a hierarchical structure with four levels: sections, divisions, groups and classes, identified by an alphabetical code or a two, three or four digit numerical code, respectively.

In this study, we considered the concordance between NACE divisions (2-digit level) and IPC codes (see \cite{Van15}, pp. 8-11, table 2), in order to analyze the decomposition of the inequality in the use frequency of patent technology codes by productive economic activities. 

Some key issues need to be addressed: 
\begin{enumerate}
\item IPC codes with subclass symbol C07B, C07C, C07F, C07G, C12S or C40B,
correspond to NACE 20 division, and IPC codes with subclass symbol C12M correspond to NACE 32 division, when they are used in a patent application together with a IPC code of the A61K008/ group. In other case, all of them correspond to NACE 21 division. So, those IPC codes can correspond to two different NACE division, meaning that they could be considered as different codes in terms of NACE correspondence.
\item There is a Co-IPC group in the list of NACE divisions for all the IPC codes not elsewhere classified. 
\item Our dataset of IPC codes, based on that correspondence with NACE divisions, can be grouped in 27 disjoint subsets whose union is equal to the entire dataset. Therefore, it gives us a partition of our dataset.
\end{enumerate}

Table \ref{tab.02} shows the total use frequency of IPC codes of each NACE division in the period 1977-2018, expressed in number of times and in percentage (as we indicated in  subsection \ref{subsection21}, in our dataset we had 72,607 different IPC codes that were used 13,615,554 times in total, in 3,786,795 patents over the period 1977-2018). It can be noted that the most popular NACE division (considering all the period 1977-2018) was NACE 26 (Manufacture of Computer, Electronic and Optical Products), with 26.28 per cent of the total activity, and the less popular NACE division was NACE 16 (Manufacture of Wood and of Products of Wood and Cork, except Furniture; Manufacture of Articles of Straw and Plaiting Materials), with only a 0.03 per cent.

For our analysis of the decomposition of the total use frequency inequality of IPC codes by productive economic activities, and based on the partition of our dataset described above, we considered the decomposition of the Theil index in two terms \cite{Sho80,Sho05}, as the sum of a within-group term $T({\bm y},n)_W$ and a between-group inequality term $T({\bm y},n)_B$,  as follows:
\begin{equation}\label{Eq.03}
T({\bm y},n)=T({\bm y},n)_W+T({\bm y},n)_B=
\sum_{g=1}^G  \displaystyle\frac{n_g \mu_g}{n \mu}T({\bm y^g},n_g)+
\sum_{g=1}^G  \displaystyle\frac{n_g \mu_g}{n \mu}
\log\left(\displaystyle\frac{\mu_g}{\mu}\right)
\end{equation}
where $G$ in the number of disjoint subsets (in our case, $G=27$ NACE divisions), 
and for each subset $g$, we have that:
${\bm y^g}=(y_1,\dots,y_{n_g})$ is the use frequency a year vector for a population of $n_g$ codes,  with mean value $\mu_g$;
$\displaystyle\frac{n_g \mu_g}{n \mu}$ is the use frequency a year share; 
and $T({\bm y^g},n_g)=
\displaystyle\frac{1}{n_g}\sum_{i=1}^{n_g}
\left(\displaystyle\frac{y_i}{\mu_g}\right)\log\left(\displaystyle\frac{y_i}{\mu_g}\right)$ 
is the Theil index of that subset. 

On the one hand, the within-group inequality component $T({\bm y},n)_W$ gives us a weighted sum of the  inequality of use frequency a year of the IPC codes of each NACE subset (a weighted sum of the Theil index $T({\bm y^g},n_g)$ of each NACE division), 
and on the other hand, the between-group inequality component  $T({\bm y},n)_B$ gives us the inequality of use frequency a year between different NACE subsets 
by replacing the use frequency of each IPC code with the mean use frequency $\mu_g$ of their respective NACE division in that year.

\begin{table}[htp]\footnotesize
\caption{Total use frequency of IPC codes of each NACE division (2-digit level). Period 1977-2018.}
{\begin{tabularx}{\textwidth}{lXrr}
\toprule
NACE & Description & \multicolumn{2}{c}{Total Use Frequency}\\[0.5ex]
Division&&(Number)&  (Percentage)\\
\midrule
10& 	Manufacture of Food Products		&111,635			&0.82\\
11& 	Manufacture of Beverages		&  14,525			&0.11\\
12& 	Manufacture of Tobacco Products	&  13,184			&0.10\\
13& 	Manufacture of Textiles			&   55,587			&0.41\\
14& 	Manufacture of Wearing Apparel 	&   18,107			&0.13\\
15& 	Manufacture of Leather and Related Products	&20,072	&0.15\\
16& 	Manufacture of Wood and of Products of
	Wood and Cork, except Furniture; Manufacture
	of Articles of Straw and Plaiting Materials		&  4,619	&0.03\\ 
17&  Manufacture of Paper and Paper Products 	&33,574	&0.25\\
18& 	Printing and Reproduction of Recorded Media	&31,817	&0.23\\
19& 	Manufacture of Coke and Refined Petroleum Products	&46,484	&0.34\\
20&  Manufacture of Chemicals and Chemical Products	&1,648,487	&12.11\\
21& 	Manufacture of Basic Pharmaceutical
	Products and Pharmaceutical Preparations	&2,383,402	&17.50\\
22& 	Manufacture of Rubber and Plastic Products	&  192,728	&1.42\\
23& 	Manufacture of Other Non-Metallic Mineral Product	&234,959	&1.73\\
24& 	Manufacture of Basic Metals		&143,390		&1.05\\
25& 	Manufacture of Fabricated Metal Products, 
	except Machinery and Equipment	&176,859		&1.30\\
26& 	Manufacture of Computer, Electronic and Optical Products
								&3,578,270	&26.28\\
27& 	Manufacture of Electrical Equipment	&880,316		&6.47\\
28& 	Manufacture of Machinery and Equipment N.E.C.&1,947,629	&14.30\\
29& 	Manufacture of Motor Vehicles,Trailers and Semi-Trailers
								&460,651		&3.38\\
30& 	Manufacture of Other Transport Equipment	&102,954		&0.76\\
31& 	Manufacture of Furniture			&40,781		&0.30\\
32& 	Other Manufacturing				&1,138,554	&8.36\\
42& 	Civil Engineering				&12,501		&0.09\\
43& 	Specialised Construction Activities	&79,148		&0.58\\
62& 	Computer Programming,
	Consultancy and Related Activities	&139,336		&1.02\\
Co\_IPC	& IPC codes not elsewhere classified&105,985	&0.78\\
		& Total	&13,615,554         	&100.00 	\\ \bottomrule
\end{tabularx}}
\label{tab.02}
\end{table}

\section{Results}
\label{sec:3}

\subsection{Total inequality in use frequency of codes}
\begin{figure}[htp]
\centering
\includegraphics[width=1\textwidth]{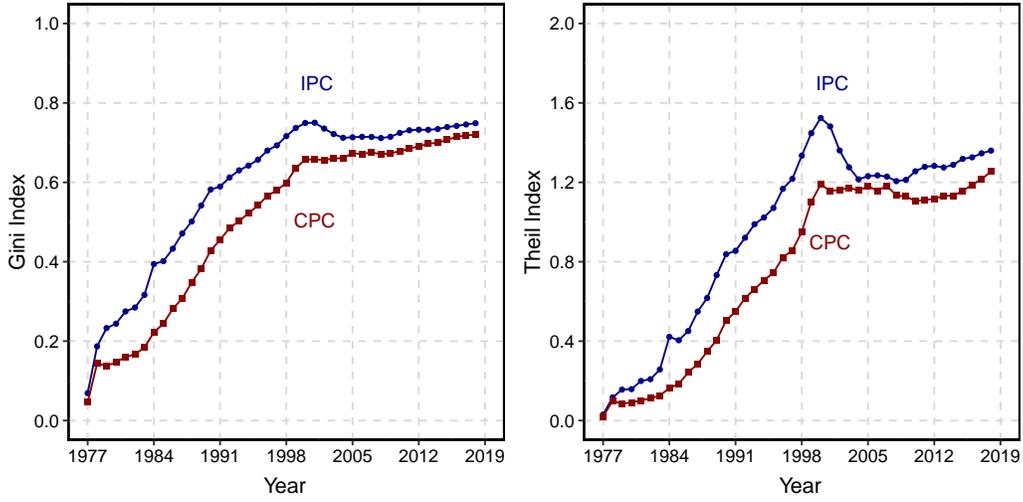}
\caption{Total inequality in use frequency of IPC and CPC codes.  Period: 1977-2018. } \label{fig.03}
\end{figure}

Figure \ref{fig.03} shows the total inequality in use frequency a year of IPC and CPC codes, measured by using the Gini index (given by Eq.(\ref{Eq.01})) and the Theil index 
(given by Eq.(\ref{Eq.02})), from 1977 to 2018. It can be observed that the Gini index reached greater values than the Theil index in all the years considered.

\begin{figure}[htp]
\centering
\includegraphics[width=1\textwidth]{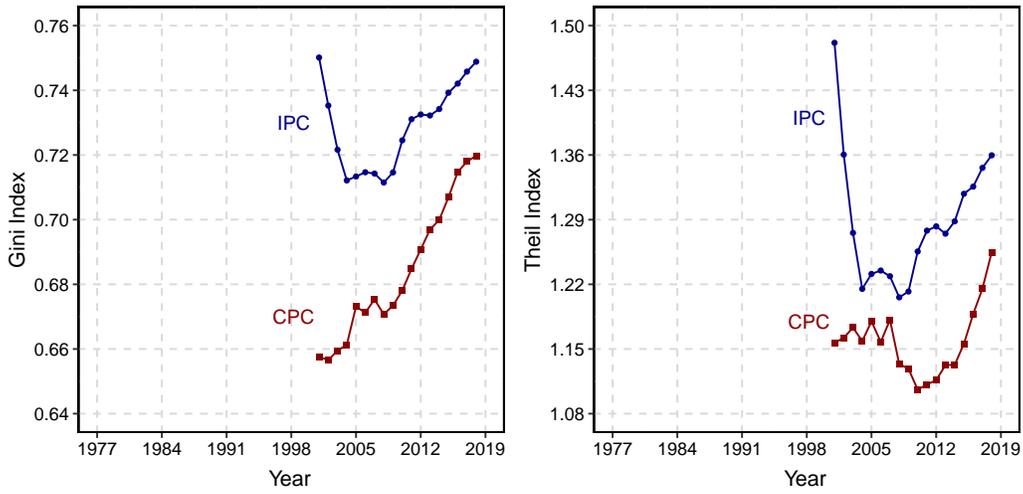}
\caption{Total inequality in use frequency of IPC and CPC codes.  Period: 2001 to 2018. } \label{fig.04}
\end{figure}

The evidence suggests that the evolution of that inequality had two clearly differentiated phases, whether we consider IPC or CPC codes, Gini or Theil index: 
\begin{itemize}
\item in the first phase, from 1977 to 2000, the total inequality showed an increasing pattern.  The Gini index increased steadily, reaching the values of 0.75 (IPC codes) and 0.66 (CPC codes) in the year 2000, and the Theil index also increased to the values of 1.52 (IPC codes) and 1.19 (CPC codes) in that year.
\item in the second phase, from 2001 to 2018, the total inequality showed a relative stabilization (in comparison with the first phase). It exhibited an U-shaped pattern in the case of IPC codes (in both Gini and Theil indexes), a slower increasing pattern in the case of the Gini index with CPC codes, and an U-shaped pattern (after some years) in the case of Theil index with CPC codes. Figure \ref{fig.04} shows a magnified image of that second phase, in which those patterns can be better observed.
\end{itemize}
Therefore, the evidence suggests that the total inequality of use frequency a year of technology codes, after those two phases, tends to be persistently high.  

\begin{figure}[htp]
\centering
\includegraphics[width=1\textwidth]{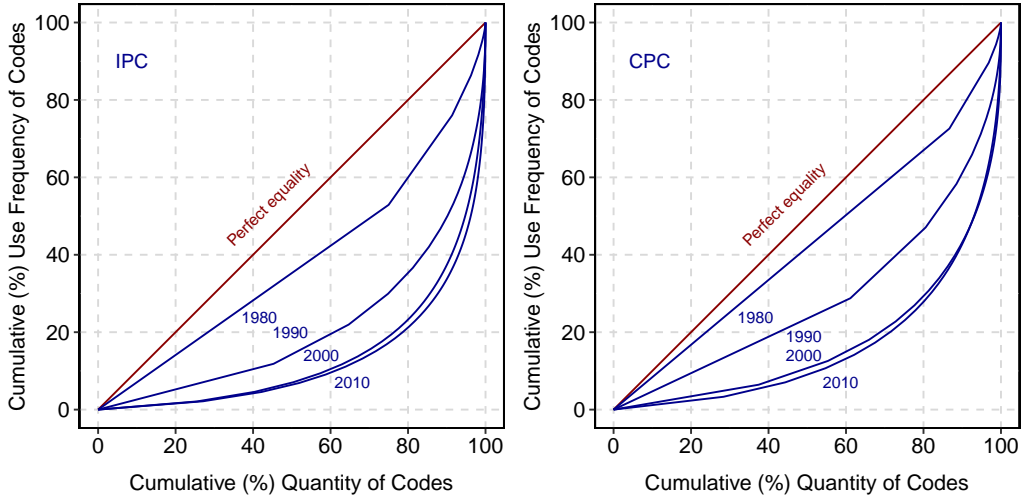}
\caption{Lorenz curves, corresponding to IPC and CPC codes, for years: 1980, 1990, 2000 and 2010.} \label{fig.05}
\end{figure}

Figure \ref{fig.05} shows the corresponding Lorenz curves, for IPC and CPC codes, obtained for four selected years: 1980, 1990, 2000 and 2010. They allows us to visualize a large increase in the inequality of use frequency of codes from 1980 to 2000 (corresponding to the first phase previously described) and a slight increase of that inequality from 2000 to 2010 (in concordance with the second phase of relative stabilization).

\subsection{Decomposition of the inequality by productive economic activities}

\begin{figure}[htp]
\centering
\includegraphics[width=1\textwidth]{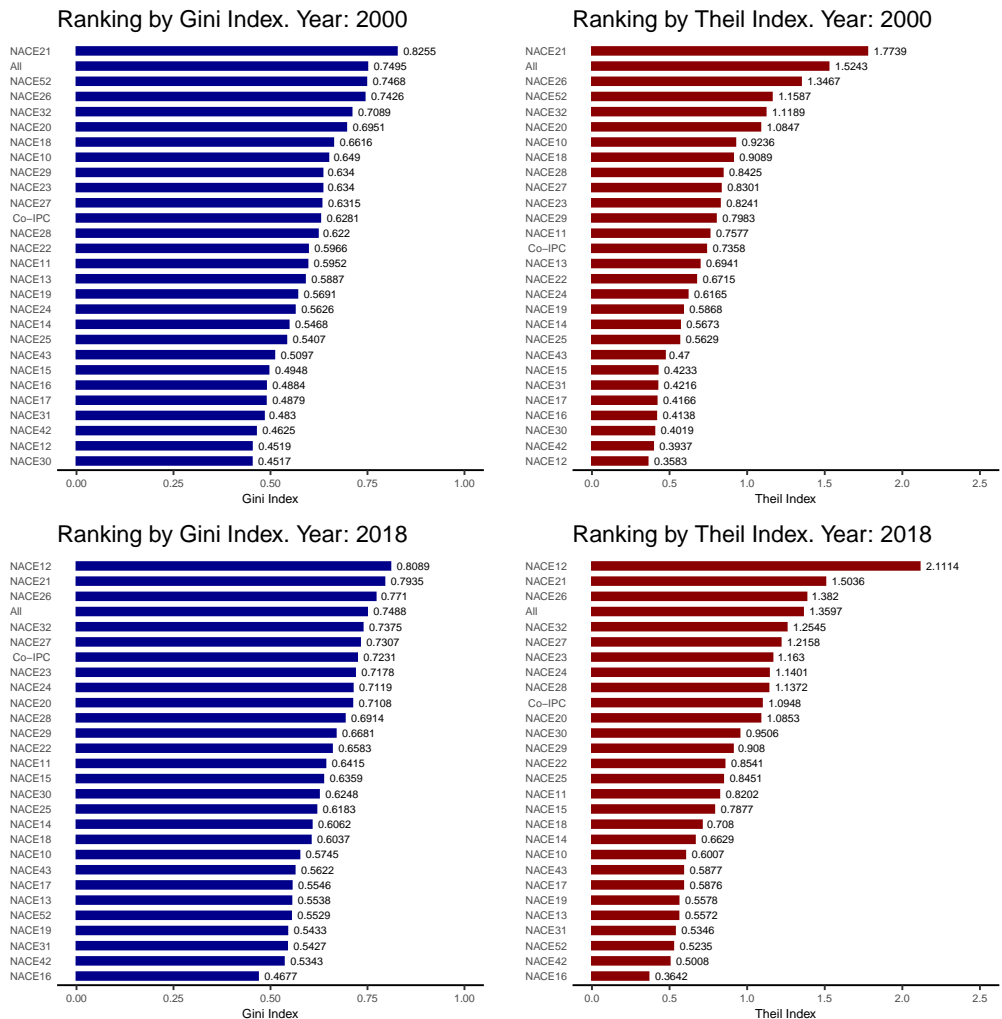}
\caption{Ranking of NACE divisions in years 2000 and 2018, in term of inequality of use frequency of IPC codes, measured by the Gini index (on the left) and by the Theil index (on the right).} \label{fig.06}
\end{figure}

Figure \ref{fig.06} shows the ranking of NACE divisions in year 2000 and year 2018 (the last year analyzed), in term of inequality of use frequency per year of IPC codes within each NACE division, measured by the Gini index (on the left) and the Theil index (on the right). 

It can be noted that, in 2018, the first three positions were occupied by NACE12 (Manufacture of Tobacco Products), NACE 21 (Manufacture of Basic Pharmaceutical Products and Pharmaceutical Preparations), and  NACE 26 (Manufacture of Computer, Electronic and Optical Products) divisions, respectively, with inequality values (from both measures) greater than the inequality value of the whole dataset (represented by the ``All" row). With respect to the last two positions, they were occupied by NACE 42 (Civil Engineering) and NACE 16 (Manufacture of Wood and of Products of Wood and Cork, except Furniture; Manufacture of Articles of Straw and Plaiting Materials) divisions. 

In comparison with the ranking in year 2000, it can be observed that some NACE divisions changed its place in the ranking significantly. For example, whereas NACE 21 division held one of the top positions in the ranking, NACE 52 division moved down to the top positions to the last positions in the ranking, and NACE 12 division did the opposite way and moved up from the last positions in 2000 to the first position in 2018.

\begin{figure}[htp]
\centering
\includegraphics[width=1\textwidth]{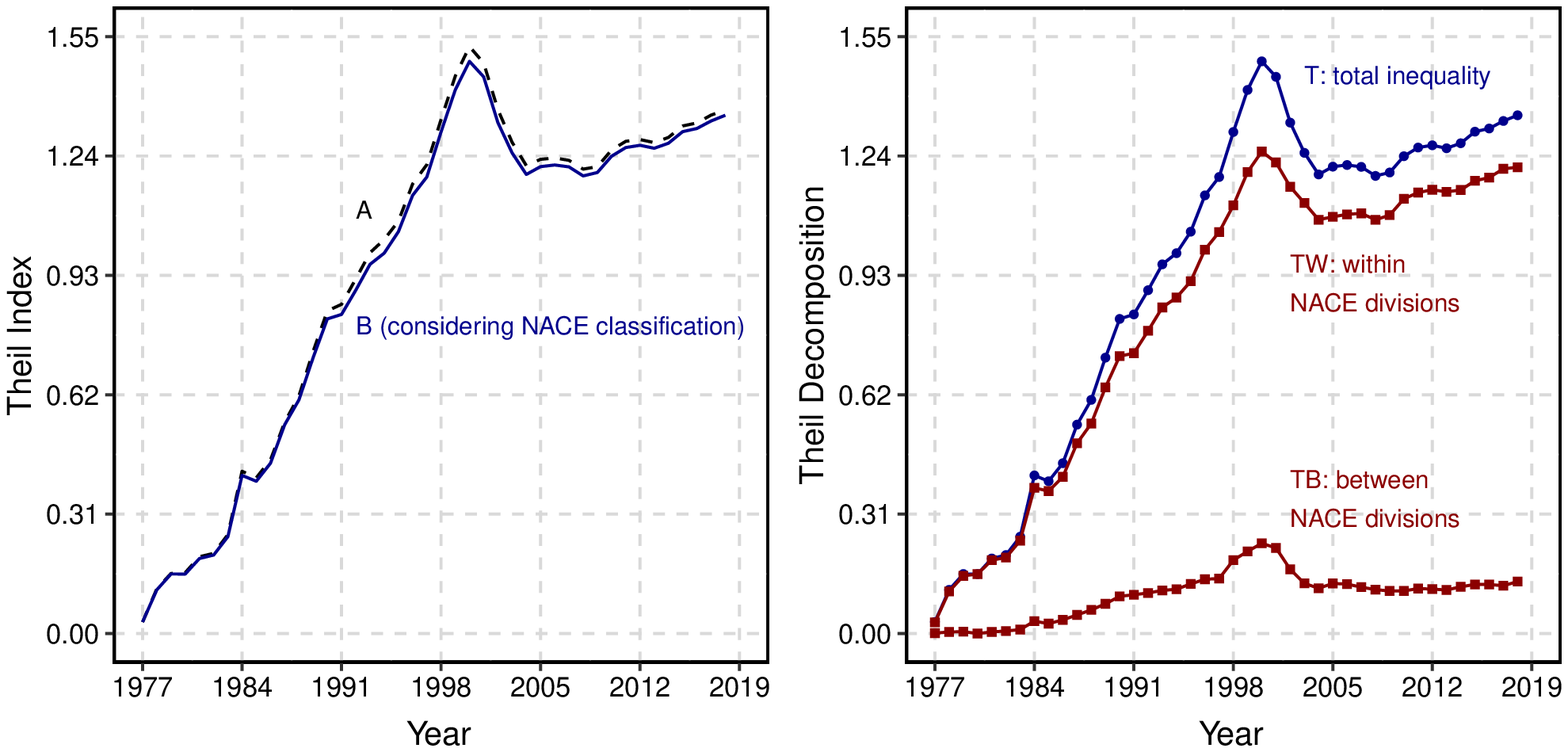}
\caption{Evolution of the within-NACE and between-NACE components, of the total inequality of use frequency per year of IPC codes, measured by the Theil index. Period: 1977-2018.} \label{fig.07}
\end{figure}

Figure \ref{fig.07} shows, on the left, the total inequality of use frequency of IPC codes from our dataset (in dashed line, line A, measured by the Theil index), in comparison with the total inequality obtained considering the different classification of the IPC codes (with subclass symbols C07B, C07C, C07F, C07G, C12M , C12S or C40B) in different NACE division, depending on whether or not they are with IPC codes of the A61K008/ group (line B, see previous section). It can be observed that lines A and B are very similar, therefore, the effect of that division is very low in the total inequality of use frequency per year obtained.

Figure \ref{fig.07} shows, on the right, the decomposition of the total inequality in two components, given by Eq.(\ref{Eq.03}): the within-group  inequality term ($TW=T({\bm y},n)_W$) and between-group inequality term ($TB=T({\bm y},n)_B$). It can be noted that the most part of the total inequality, in all the years considered, was due to the within-group component. 

Therefore, the evidence suggests that the total level of inequality of use frequency of IPC codes was mainly driven by the differences in the use frequency of those codes within each NACE division. And that the contribution of the differences in average use frequency between NACE divisions (taking into account the population of each one) was very low.

\section{Conclusions}
\label{sec:4}

This paper measures for the first time, and as far as we know, the inequality in the use frequency of patent technology codes. For that, we selected the patent applications filed under the Patent Co-operation Treaty (PCT) at international phase, with the European Patent Office (EPO) as designated office, and considered the information of IPC and CPC codes from those patents. First, we measured that inequality by using the Gini index and the Theil index, and finally, we 
studied the decomposition of that inequality by productive economic activities.

We found that total inequality in use frequency a year of technology codes had two clearly differentiated phases. During the first phase (1977-2000), the total inequality showed an increasing pattern. In the second phase (2001-2018), the total inequality showed a relative stabilization. 

We found that total inequality in use frequency a year of technology codes tends to be persistently high, whether we consider IPC or CPC codes, or whether we measure that inequality using the Gini index or the Theil index.

After grouping IPC codes by productive economic activities, we found that total inequality is mainly driven by the differences in the use frequency of those codes within each group, with a low contribution of the differences between groups. 

As future research, this study can be extended to other patent databases, for example, 
USPTO (US Patent and Trademark Office) or JPO (Japanese Patent Office) databases.
Other open challenges are to analyze the factors related with those two phases of inequality growth or with the changes in the inequality of use frequency of codes within NACE divisions.

\bibliographystyle{plain}

\begin{thebibliography}{11}
%
\bibitem{You15}
Youn, H., Strumsky, D., Bettencourt, L. M., \& Lobo, J. (2015). Invention as a combinatorial process: evidence from US patents. \emph{Journal of the Royal Society Interface}, 12(106), 20150272.

\bibitem{Str10}
Strumsky, D., Lobo, J., \& Tainter, J. A. (2010). Complexity and the productivity of innovation. \emph{Systems Research and Behavioral Science}, 27(5), 496-509.

\bibitem{Str12}
Strumsky, D., Lobo, J., \& Van der Leeuw, S. (2012). Using patent technology codes to study technological change. \emph{Economics of Innovation and New 
technology}, 21(3), 267-286.

\bibitem{Str15}
Strumsky, D., \& Lobo, J. (2015). Identifying the sources of technological novelty in the process of invention. \emph{Research Policy}, 44(8), 1445-1461.

\bibitem{Cha17}
Chang, S. H. (2017). The technology networks and development trends of university-industry collaborative patents. \emph{Technological Forecasting and Social 
Change}, 118, 107-113.

\bibitem{Cha19}
Chae, S., \& Gim. J. (2019). A study on trend analysis of applicants based on patent classification systems. \emph{Information}, 10(12), 364.

\bibitem{Gol20}
Goldschlag, N., Lybbert, T. J., \& Zolas, N. J. (2020). Tracking the technological composition of industries with algorithmic patent concordances. \emph{Economics of Innovation and New Technology}, 29(6), 582-602.

\bibitem{Wei98}
Weitzman, M. L. (1998). Recombinant growth. \emph{The Quarterly Journal of Economics}, 113(2), 331-360.

\bibitem{Art09}
Arthur, W. B. (2009). \emph{The nature of technology: What it is and how it evolves}. Simon and Schuster.

\bibitem{Hua16}
Huang, J. Y. (2016). Patent portfolio analysis of the cloud computing 
industry. \emph{Journal of Engineering and Technology Management}, 39, 45-64.

\bibitem{Smo16}
Smojver, V., \v{S}torga, M., \& Poto\v{c}ki, E. (2016). An extended methodology for the assessment of technical invention evolution. In DS 84: Proceedings of the DESIGN 2016 14th International Design Conference, 1135-1144.

\bibitem{Har19}
Harrigan, K. R., \& Fang, Y. (2019). Financial implications of technology-class code popularity and usage among industry competitors. \emph{Scientometrics}, 121(1), 25-51.

\bibitem{Sen97}
Sen, A. K. (1997). From income inequality to economic inequality. \emph{Southern Economic Journal}, 64(2), 384-401.

\bibitem{Pik03}
Piketty, T., \& Saez, E. (2003). Income inequality in the United States, 
1913-1998. \emph{The Quarterly Journal of Economics}, 118(1), 1-41.

\bibitem{Sar19}
Sarabia, J. M., Jord\'a, V., \& Prieto, F. (2019). On a new Pareto-type distribution with applications in the study of income inequality and risk analysis. \emph{Physica A: Statistical Mechanics and its Applications}, 527, 121277.

\bibitem{Cag08}
Cagetti, M., \& De Nardi, M. (2008). Wealth inequality: Data and 
models. \emph{Macroeconomic Dynamics}, 12(S2), 285-313.

\bibitem{Zuc19}
Zucman, G. (2019). Global wealth inequality. \emph{Annual Review of 
Economics}, 11, 109-138.

\bibitem{Til87}
Tilak, J. B. (1987). \emph{The economics of inequality in education}. Sage Publications India (Pvt) Ltd.

\bibitem{Jor17}
Jord\'a, V., \& Alonso, J. M. (2017). New estimates on educational attainment using a continuous approach (1970-2010). \emph{World Development}, 90, 281-293.

\bibitem{Coc20}
Coco, G., Lagravinese, R., \& Resce, G. (2020). Beyond the weights: a multicriteria approach to evaluate inequality in education. \emph{The Journal of Economic Inequality}, 18(4), 469-489.

\bibitem{Ezc07}
Ezcurra, R., Pascual, P., \& Rap\'un, M. (2007). Spatial inequality in productivity in the European union: sectoral and regional factors. \emph{International Regional Science Review}, 30(4), 384-407.

\bibitem{Adr19}
Adri\'an Risso, W., \& S\'anchez Carrera, E. J. (2019). On the impact of innovation and inequality in economic growth. \emph{Economics of Innovation and New 
Technology}, 28(1), 64-81.

\bibitem{Mar05}
Marsili, O., \& Salter, A. (2005). Inequality of innovation: skewed distributions and the returns to innovation in Dutch manufacturing. \emph{Economics of Innovation and New Technology}, 14(1-2), 83-102.

\bibitem{OHu11}
\'O Huallach\'ain, B., \&  Lee, D. S. (2011). Technological specialization and variety in urban invention. \emph{Regional Studies}, 45(1), 67-88.

\bibitem{Man21}
Mancusi, M. L. (2021). Geographical concentration and the dynamics of countries' specialization in technologies. \emph{Economics of Innovation and New 
Technology}, 12(3), 269-291.

\bibitem{Biu20}
Biurrun, A. (2020). New evidence toward solving the puzzle of innovation and 
inequality. The role of institutions. \emph{Economics of Innovation and New Technology}, 1-22.

\bibitem{Mar08}
Maraut, S., Dernis, H., Webb, C., Spiezia, V., \& Guellec, D. (2008). \emph{The OECD REGPAT database: a presentation}.

\bibitem{Oec21}
OECD, REGPAT database, July 2021.\\ 
{\footnotesize https://www.oecd.org/sti/inno/intellectual-property-statistics-and-analysis.htm}

\bibitem{Oec09}
OECD Patent Statistics Manual. 2009. OECD, Paris.

\bibitem{Arn18}
Arnold, B. C., \& Sarabia, J. M. (2018). Inequality Measures. In \emph{Majorization and the Lorenz Order with Applications in Applied Mathematics and Economics}, 45-114. Springer, Cham.

\bibitem{Bro94}
Brown, M. C. (1994). Using Gini-style indices to evaluate the spatial patterns of health practitioners: theoretical considerations and an application based on Alberta 
data. \emph{Social Science} \& \emph{Medicine}, 38(9), 1243-1256.

\bibitem{All78}
Allison, P. D. (1978). Measures of inequality. \emph{American Sociological Review}, 865-880.

\bibitem{Sho80}
Shorrocks, A. F. (1980). The class of additively decomposable inequality measures. \emph{Econometrica: Journal of the Econometric Society}, 48(3), 613-625.

\bibitem{Eur08}
Eurostat methodologies and working papers. (2008). \emph{NACE Rev. 2. Statistical classification of economic activities in the european community}. Publications Office of the European Union: Luxembourg.

\bibitem{Van15}
Van Looy, B., Vereyen, C., \& Schmoch, U. (2015). \emph{Patent Statistics: Concordance IPC V8-NACE Rev. 2 (version 2.0)}. EUROSTAT, Luxembourg.

\bibitem{Sho05}
Shorrocks, A., \& Wan, H. (2005). Spatial decomposition of inequality. \emph{Journal of Economic Geography}, 5(1), 59-81.

\end{thebibliography}

\end{document}